\documentclass[aps,twocolumn,showpacs,showkeys,letterpaper]{revtex4}
\usepackage{graphicx}
\begin{document}

\title{No singularities at the phase transition in the Dicke model}

\author{O. Casta\~nos, E. Nahmad-Achar, R. L\'opez-Pe\~na, and J. G. Hirsch}

\affiliation{Instituto de Ciencias Nucleares,
Universidad Nacional Aut\'onoma de M\'exico \\
Apdo. Postal 70-543, Mexico D. F., C.P. 04510}

\date{\today}

\begin{abstract}
The Dicke Hamiltonian describes the simplest quantum system with atoms interacting with photons: N two level atoms inside a perfectly reflecting cavity which allows only one electromagnetic mode. It has also been successfully employed to describe superconducting circuits which behave as artificial atoms coupled to a resonator. The system exhibits a transition to a superradiant phase at zero temperature. When the interaction strength reaches its critical value, both the number of photons and of atoms in excited states in the cavity, together with their fluctuations, exhibit a sudden increase from zero. Employing symmetry-adapted coherent states it is shown that these properties scale with the number of atoms, that their reported divergences at the critical point represent the limit when this number goes to infinity, and that in this limit they remain divergent in the superradiant phase. Analytical expressions are presented for all observables of interest, for any number of atoms. Comparisons with exact numerical solutions strongly support the results.
\end{abstract}


\maketitle

The quantum description of a system of $N$ non-interacting two-level atoms under the action of a one mode electromagnetic field allowed Dicke~\cite{dicke} to predict the superradiance phenomenon, where all atoms absorb and emit light collectively and coherently \cite{hepp}. For N=1 (the Jaynes-Cummings model), the system exhibits quantum features such as collapse and revivals of the atomic inversion~\cite{eberly} and squeezing of the radiation field~\cite{kulinski}.   

The Dicke quantum phase transition has recently been observed employing a superfluid gas in an optical cavity \cite{baumann}. While this breakthrough was achieved using time dependent fields dressing the system, it has been argued that superconducting circuit QED systems will exhibit a stationary superradiant phase \cite{Nat10}. The description of these systems is given in terms of the same Dicke Hamiltonian \cite{Dim07}, with minor modifications \cite{nagy}. The theoretical description is obtained by mapping the original Hamiltonian to a system of two bosons, making a Holstein-Primakoff realization of the quasi-spin operators which represent the collective atomic excitation, and truncating the infinite series up to first or second order assuming that the expectation value of the number of excited atoms (or the number of photons) is small compared with the total number of atoms N \cite{Ema03,Dim07,nagy,Nat10}. This system exhibits a transition to a superradiant phase at zero temperature. When the interaction strength reaches its critical value, a divergence is predicted in the expectation value of the number of photons \cite{Dim07,nagy}, the number of atoms in excited states \cite{nagy} and their fluctuations \cite{Ema03,Dim07}.

In this contribution we show, employing symmetry adapted coherent states, that these properties scale with the number of atoms, that their reported divergences at the critical point represent the limit when this number goes to infinity, and that in this limit they remain divergent in the superradiant phase, contrary to what is stated in some of the references mentioned above \cite{Ema03,Dim07,nagy,Nat10}. 
The ground state of the Dicke model is obtained using the tensorial product of spin coherent and Weyl states as a trial wavefunction. The Hamiltonian expectation value is calculated, and its critical points found, showing the existence of the quantum phase transition. The superradiant phase is described employing combinations of coherent states which preserve the Hamiltonian $C_2$ symmetry \cite{Cas10}. With them, analytical expressions are obtained for all observables of interest, which are valid for any number of atoms. Comparisons with exact numerical solutions strongly support the results.

The Dicke Hamiltonian involves the collective interaction of $N$ two-level atoms with energy separation equal to $\hbar \tilde\omega_A$ with a one mode radiation field of frequency $\tilde\omega_F$. Its deduction from the quantization of the electromagnetic field makes the following assumptions: (i) The dipole approximation or the long wavelenght limit in such a way that the electric field is evaluated at the center of mass of the atoms and is independent of the position. (ii) Only the two atomic levels are interacting with the electromagnetic field. The lowest energy state is at least metastable and thus the decay to other energy levels can be neglected.   
The Dicke model Hamiltonian for $N$ identical $2$-level atoms immersed in a one mode electromagnetic field, is given by
	\begin{equation}
		H_{D}= a^{\dagger}a + \omega_{A}\,J_{z}
		+\frac{\gamma}{\sqrt{N}}\left(a^{\dagger} + a \right) \left( \,J_{-}
		+ \,J_{+}\right)\ ,
		\label{D}
	\end{equation}
where $\omega_A= \tilde{\omega}_A/\tilde{\omega}_F \geq 0$ is given in units of the frequency of the field, and $\gamma= \tilde{\gamma}/\tilde{\omega}_F$ is the (adimensional) coupling parameter. The operators $a$, $a^{\dagger}$, denote the one-mode annihilation and creation photon operators, respectively, $J_{z}$ the atomic relative population operator, and $J_{\pm}$ the atomic transition operators.

The Dicke Hamiltonian has a dynamical symmetry associated with the projectors of the symmetric and antisymmetric representations of the cyclic group $C_2$, given by, 
	\[
	P_S= \frac{1}{2} \bigl( 1 + e^{i \pi \hat{\Lambda}} \bigr) \, , 
	\qquad P_A= \frac{1}{2} \bigl( 1 - e^{i \pi \hat{\Lambda}} \bigr) \, .
	\]
This symmetry allows the classification of the eigenstates in terms of the parity of the eigenvalues $\lambda = j +m + \nu$ of the excitation number operator 
	\[
	\hat\Lambda = \sqrt{\hat{J}^2 + 1/4} -1/2 + \hat{J}_z + a^{\dagger}a ,
	\]
with $j=N/2$, $j+m$ the number of atoms in their excited state, and $\nu$ the number of photons.

To obtain analytical expressions the variational procedure described in~\cite{ocasta1} is employed. We use as a trial state the direct product
of coherent states in each subspace: Heisenberg-Weyl states $\vert\alpha\rangle$ for the photon
sector~\cite{meystre,manko} and $SU(2)$ or spin states $\vert\zeta\rangle$ for the particle sector~\cite{gilmore1972,Cas05}, i.e., $\vert\alpha, \,\zeta\rangle=\vert\alpha\rangle \otimes\vert\zeta\rangle$. Due to the symmetry properties of the Dicke Hamiltonian, it is convenient to replace $m$ by $\lambda- j-\nu$:
 Explicitly, it is given by the expression
	\begin{eqnarray}
		\vert\alpha ,\,\zeta\rangle =&
		{\cal N}_{coh}^{-1/2}
		\sum\limits_{\lambda=0}^{\infty}
		\sum\limits_{\nu=\max[0,\,\lambda-2 j]}^{\lambda}\,
		\frac{\alpha^{\nu}}{\sqrt{\nu!}}\,
		\left( \begin{array}{c} 2j 
		\\ \lambda - \nu \end{array} \right)^{1/2} \times \nonumber	\\
		& \zeta^{\lambda - \nu}\,\vert\nu\rangle \otimes \vert j,\,\lambda - \nu - j\rangle \ ,
	\label{trial}
	\end{eqnarray}
where the ket $\vert \nu \rangle$ is an eigenstate of the photon number operator, $\vert j,\, m \rangle$ is a Dicke state with $j$ denoting the eigenvalue of $J^2$, and $m$ the corresponding eigenvalue of $J_z$. The trial state contains $N=2j$ particles distributed in all the possible ways between the two levels and up to an infinite number of photons in the cavity. 
The normalization is 
\begin{equation}
{\cal N}_{coh} = \left(1+\left|\zeta\right|^{2}\right)^{N} \exp\left|\alpha\right|^{2}
\label{norm_coh}
\end{equation}

The expectation value of the Dicke Hamiltonian with respect to $\vert\alpha ,\,\zeta\rangle$, when
divided by the total number of particles, is called the energy surface of the Dicke Hamiltonian. It is a function of four variables and two parameters,
	\begin{eqnarray}
		E(q,\,p,\,\theta,\,\phi)= 
\frac{1}{2 N} \left( p^2 + q^2 \right) - \frac{1}{2 } \omega_A \cos\theta 
\nonumber \\
+ \frac{\sqrt 2 \gamma }{\sqrt N} q \, \sin\theta \cos\phi \ ,
	\end{eqnarray}
where use has been made of the harmonic oscillator realization of the field, and of the stereographic projection for the quasi-spin sector
	\begin{equation}
		\alpha=\frac{1}{\sqrt{2}}\left(q+i\,p\right)\ ,\qquad
		\zeta = \tan\left(\frac{\theta}{2}\right)\,\exp\left(-i\,\phi\right)\ ,
		\label{dzeta}
	\end{equation}
$(q,p)$ correspond to the expectation values of the quadratures of the field and $(\theta,\phi)$ determine a point on the Bloch sphere.

The minimum and degenerate critical points are obtained by means of the catastrophe formalism~\cite{gilmore3}. The minimum critical points are, 
for $\omega_{A} \geq 4\, \gamma^{\, 2}$: $\, \theta_{c}=0\,$, $q_{c}=0\, $, and $p_{c}=0 \,$;
and for $ 0 \leq \omega_{A} < 4\, \gamma^{\, 2} $ : 
$\theta_{c}=\arccos\left(\frac{\omega_{A}}{4 \, \gamma^{\, 2}}\right)$,
$\phi_c=0$,
$q_{c}=- \, \sqrt{2 N} \, \gamma \, \sqrt{1- \frac{\omega_A^{\, 2}}{16 \gamma^{\, 4}}} \cos{\phi_c} \,$,  
$p_{c}=0 \,$.
The first group contains the minimum critical points in the normal phase, the second one describes the minimum critical points in the superradiant phase. The phase transition occurs at 
$\omega_{A} = 4\, \gamma_c^{2}$, i.e. $\gamma_c = \sqrt{\omega_{A}}/2$. In dimensional units, 
$\tilde{\gamma}_c = \sqrt{\tilde{\omega}_{A} \tilde{\omega}_F}/2$, in agreement with Ref. \cite{Ema03,nagy}.

 The associated energies, expected number of photons $ \langle \hat{N}_{ph}\rangle = \vert \alpha_c \vert^2$ and of excited atoms $\langle  n_e \rangle =\langle {\frac N 2} + J_z \rangle = {\frac N 2} (1 - \, \hbox{cos} \theta_{c})$ are  
\begin{eqnarray}
		 &E_{normal} = -2 N\, \gamma_c^2 \, , 
&E_{super} = - N \,\gamma^2 \left( \left(\frac{\gamma_c}{\gamma}\right)^4 +1 \right) \, .
\nonumber	 \\
&\langle \hat{N}_{ph}\rangle_{normal} = 0 \, , 
&\langle \hat{N}_{ph}\rangle_{super} = 
N  \, \gamma^2 \, \left(1- \left(\frac{\gamma_c}{\gamma}\right)^4  \right), \nonumber \\
&\langle  n_e\rangle_{normal} = 0  \, , 
&\langle  n_e \rangle_{super} =  {\frac N 2} \left (1- \, \left(\frac{\gamma_c}{\gamma}\right)^2 \right) .
\label{ecoherente}
\end{eqnarray}
These mean field expressions exactly coincide with those presented in \cite{Ema03}, and reproduce with precision the numerical results even for relatively small number of particles. The mean field results fail badly, however, in the description of the fluctuations of these observables, a limitation also found in the Tavis-Cummings model, where the counter-rotating terms are absent \cite{ocasta2}. The reason is that in the superradiant phase the state $ \vert\alpha ,\,\zeta\rangle$ breaks the $C_2$ symmetry of the Hamiltonian, mixing all $\lambda$ values, and having an expectation value of $\langle \hat{q} \rangle =  q_{c}$ different from zero, implying that $(\Delta q)^2 = 1/2$.

In order to go beyond this mean field treatment, to obtain a symmetry-adapted description of the ground state of the Dicke model in the superradiant phase, which would preserve the $C_2$ symmetry of the Hamiltonian, we perform a projection over the $\lambda$'s in Eq. (\ref{trial}), which is equivalent to requiring that the sum  be carried over $\lambda$ even or odd. This represents a generalization of the restoration of the Hamiltonian symmetry in the LMG \cite{Cas06} and Tavis-Cummings Hamiltonians \cite{ocasta2}.  The two resulting orthogonal states with $\lambda$ even or odd $ \vert\alpha_c ,\,\zeta_c, \pm \rangle$ are even or odd under $e^{i \pi \hat{\Lambda}}$. 
Their normalizations are:
\begin{equation}
{\cal N_{\pm}} = 
	 \frac{2^{N-1}}{\left(1 + \left(\frac{\gamma_c}{\gamma}\right)^2 \right)^{N}} 
	\bigl( 1 \pm F(N, \gamma, \gamma_c) \bigr) \, .
\label{norm_eo}
\end{equation}
where we have defined
	\begin{equation}
	F(N, \gamma, \gamma_c) = \biggl(\frac{\gamma_c}{\gamma}\biggr)^{ 2 N} \, 
	\exp{\left\{ 
	- 2 N \gamma^2 \left( 1  - \left(\frac{\gamma_c}{\gamma}\right)^{\!\!4} \right)  
	\right\} } \, .
	\end{equation}

From them analytical expressions can be found for all the observables of interest \cite{Cas10}. Those for the even and odd energy surfaces are
	\begin{equation}
	\frac{\langle H \rangle_\pm}{N}  =  
	\gamma^2 \left(-3 + \left(\frac{\gamma_c}{\gamma}\right)^4 
	+  \frac{2 }{1 \pm F } 
	\left(1 -  \left(\frac{\gamma_c}{\gamma}\right)^{\!\!4} \right)  \right)	\, .
	\end{equation}
It should be noted that $F(N, \gamma, \gamma_c)\rightarrow 0 $ when $N\rightarrow \infty$ very rapidly. For example, $F(100, 1.03 \gamma_c , \gamma_c) < 10^{-5}$. It implies that the two states with different parity $C_2$ become degenerate. Notice that the origin of this degeneracy is the restoration of the $C_2$ symmetry, not its breaking.

The expectation value of the population of atoms in the two level system is
	\begin{equation}
 	\frac{\langle \hat{J}_z \rangle_{\pm}}{N}   =  - {\frac 1 2}  \left(\frac{\gamma}{\gamma_c}\right)^2
	\Biggl( 1 - \frac{1 - \left(\frac{\gamma_c}{\gamma}\right)^{\!\!4} }{1 \pm F} \Biggr)  \, ,
	\end{equation}
and the expectation value of the number of photons is given by
	\begin{equation}
	\frac{\langle \hat{N}_{ph} \rangle_{\pm}}{N}   =  \gamma^2 \, 
	\left(1 -  \left(\frac{\gamma_c}{\gamma}\right)^{\!\!4} \right)
	\frac{ 1 \mp  F} { 1 \pm F}  \, .
	\end{equation}
When $F \rightarrow 0$ both expressions coincide with the mean field ones, given in Eqs. (\ref{ecoherente}), as expected. 

The quadrature components of the electromagnetic field change the value of $\lambda$ in one unit, so their expectation values with respect to the projected states are equal to zero. For the same reason the expectation values of matter observables $\langle J_{x}\rangle $ and$ \langle J_{y}\rangle$ are null. The symmetry projection makes all the difference with respect to the mean field estimations, where these expectation values grow with the number of particles.

The expectation values of squared quadratures of the one-mode electromagnetic field are given by 
	\begin{eqnarray}
	\frac{\langle \hat{q}^2 \rangle_{\pm}}{N}   &=&  \frac{1}{2 \, N} \, +  2 \gamma^2 \left(1 -  \left(\frac{\gamma_c}{\gamma}\right)^{\!\!4} \right)
	\frac{1}{1 \pm F}   \, , \label{deltaq2} \nonumber \\
	\frac{\langle \hat{p}^2 \rangle_{\pm}}{N}   &=&  \frac{1}{2 \, N} \, -  2 \gamma^2 \left(1 -  \left(\frac{\gamma_c}{\gamma}\right)^{\!\!4} \right)
	\frac{F}{F \pm 1}   \, . 
	\end{eqnarray}

The expectation value of the operators $\hat{J}_x^2$ and $\hat{J}_y^2$ are
	\begin{eqnarray}
	\frac{\langle \hat{J}_x^2 \rangle_\pm}{N^2} &=& \frac{1}{4 N} \Bigg\{ 1 +  
	\left(1 -  \left(\frac{\gamma_c}{\gamma}\right)^{\!\!4} \right)
	\frac{ ( N -1) }{1 \pm F} \Bigg\}
	\, , \nonumber \\
	\frac{\langle \hat{J}_y^2 \rangle_\pm}{N^2} &=& \frac{1}{4 N} \Bigg\{ 1 + 
	\left(1 -  \left(\frac{\gamma_c}{\gamma}\right)^{\!\!-4} \right)
	\frac{ ( N -1) F }{F \pm 1} 		\Bigg\} 
	 \, . \label{deltaJ2}
	\end{eqnarray}
\begin{figure}[h]
\scalebox{0.8}{\includegraphics{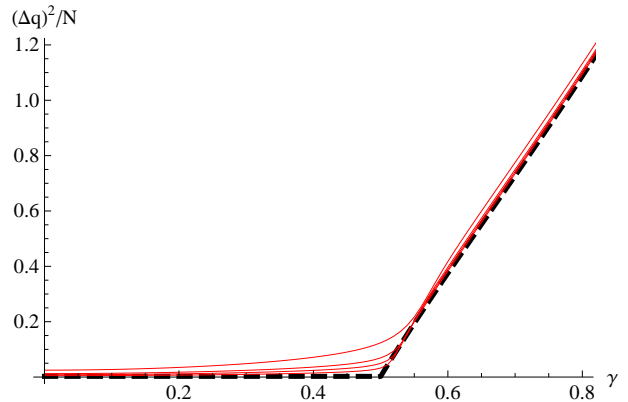}}\qquad\qquad
\caption{\label{deltaq}
Squared fluctuations of the first photon quadrature $(\Delta q)^2/N$ at resonance, $\omega_A = 1$, calculated through numerical diagonalization of the matrix Hamiltonian for $N= 10,\, 20,\, 30$ and $50$ (red thin lines ) and from Eq. 12 (black thick dashed line).  As $N$ increases the exact quantum dispersion approaches the calculated through adapted states: the upper thin curve corresponds to $N=10$, while the lower thin line corresponds to $N=50$. All the quantities plotted are dimensionless. (Color online) }
\end{figure}

\begin{figure}[h]
\scalebox{0.8}{\includegraphics{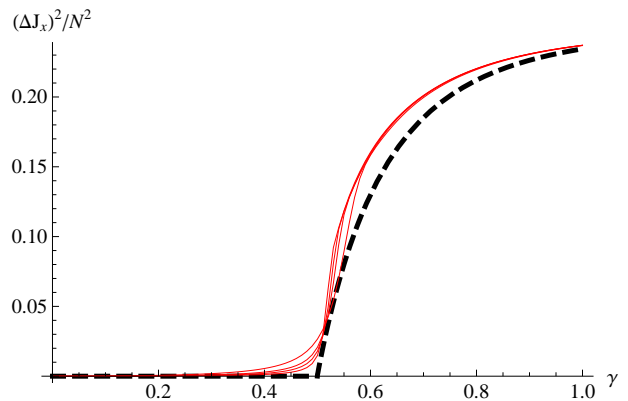}}\qquad\qquad
\caption{\label{deltaj}
Squared fluctuations of the atomic transition operator $(\Delta J_x)^2/N^2$ at resonance, $\omega_A = 1$, calculated through numerical diagonalization of the matrix Hamiltonian for $N= 10,\, 20,\, 30$ and $50$ (red thin lines) and from Eq. 13 (black thick dashed line).  As $N$ increases the exact quantum dispersion approaches the calculated through adapted states: the upper thin curve corresponds to $N=10$, while the lower thin line corresponds to $N=50$. All the quantities plotted are dimensionless.(Color online)  }
\end{figure}

In Fig. \ref{deltaq} the squared fluctuations of the quadrature $(\Delta q)^2$ at resonance ($\omega_A=1$) are presented, divided by the total number of particles $N$, as a function of the coupling strength $\gamma$. The thin red lines were calculated by numerical diagonalization of the Hamiltonian for N= 20, 40, 60 and 100, while the thick green line was obtained employing Eq. (\ref{deltaq2}). It is clear that the fluctuations scale with the number of particles. 

In Fig. \ref{deltaj} the squared fluctuations of the atomic observable $(\Delta J_x)^2$ at resonance ($\omega_A=1$) are presented, divided by the square of the total number of particles $N$, as a function of the coupling strength $\gamma$. The thin red lines were calculated by numerical diagonalization of the Hamiltonian for N=  20, 40, 60 and 100, while the thick green line was obtained employing Eq. (\ref{deltaJ2}). In this case the fluctuations scale with the square of the number of particles.

It is worth to study the superradiant phase in the limit $\gamma_c / \gamma \rightarrow 0$, which is exactly solvable \cite{Ema03a,Chen08}.
The expectation values discussed above become:
$\langle \hat{N}_{ph}\rangle \rightarrow N \,\gamma^2, \quad
\langle n_e \rangle \rightarrow N/2, \quad
(\Delta q)^2 \rightarrow  2 N \,\gamma^2 +\frac{1}{2},   \quad 
(\Delta p)^2  \rightarrow \frac{1}{2}, \quad
\xi_x^2 = 4 (\Delta J_x)^2/N  \rightarrow N, \quad
 \xi_y^2 =4 (\Delta J_y)^2/N \rightarrow 1.
$

In this limit both the average number of photons and the fluctuations in the $q$ quadrature scale with the number of particles N and with $\gamma^2$. The average number of excited atoms goes to $N/2$, and with it $\langle n_e \rangle /N \rightarrow 1/2$. It clarifies why any Taylor expansion of this quantity around zero, truncated to second or third order, cannot reproduce the asymptotic behavior.

The squeezing in spin observables is given by $\xi_i = \sqrt{4 (\Delta J_i)^2/N}$, where $J_i$ are the  perpendicular components to the average angular momentum vector $\langle \vec J \rangle$ \cite{Kita93}. There is squeezing when $\xi_i <1$. For the Dicke model $\langle \vec J \rangle = \langle J_z \rangle$ \^{k}, and the perpendicular components of the angular momentum are $J_x$ and $J_y$. The squared squeezing observables $\xi_i^2$ are, in the limit $\langle n_e \rangle /N << 1$, very close to the fluctuations of the quadratures of the atomic degrees of freedom studied in \cite{Ema03}. In a similar fashion as the number of excited atoms, the squeezing parameter $\xi_x^2$ scales with the number of atoms.

It is worth mentioning that Eqs. (\ref{deltaq2},\ref{deltaJ2}) show that $(\Delta p)^2$ and $\xi_y$ have some squeezing around $\gamma_c$, as found in \cite{Ema03}. This is an effect of order 1/N, which becomes negligibly small when the number of atoms grows. The excitation energy of the first excited state is also of this order. These aspects will be analyzed in detail in a forthcoming publication.

What is the origin of the divergences in the number of photons and its fluctuation at the critical value of $\gamma$ reported in \cite{Ema03,Dim07,nagy}? It is found in the normal phase when the limit $N\rightarrow \infty$ is assumed, obtaining a Hamiltonian bilinear in two kinds of bosons without memory of the number of particles, except for the ground state energy which has a constant displacement $-N \,\omega_A / 2$. A Bogoliubov transformation allows the exact diagonalization of this Hamiltonian, and the expectation values of the number of photons, the number of atoms in excited states, and their fluctuations, all diverge at $\gamma_c$. It agrees with the exact results reported above when the number of atoms N goes to infinity. But, in contrast to what is found in  \cite{Ema03,Dim07,nagy}, we have shown that at the superradiant phase they will remain divergent with N for any $\gamma > \gamma_c$. This finding relies on the use of symmetry projected SU(2)$\times$U(1) coherent states. The expectation values obtained employing a truncated Hamiltonian completely disagree with the numerically exact ones at the superradiant phase because the assumption $\langle n_e\rangle << N$ is not valid for $\gamma >> \gamma_c$, and the $C_2$ symmetry of the Hamiltonian is not preserved. 

In summary, we have presented analytic expressions for the ground and first excited states of the Dicke Hamiltonian in the superradiant phase which have the appropriate $C_2$ symmetry. From them, analytical expressions for all expectation values of interest were obtained, valid for any number of atoms.
The procedure was the following: the expectation value of the Hamiltonian with respect to the tensorial product of Weyl and SU(2) coherent states defines an energy surface depending on phase space variables and parameters, whose critical points were determined. The minimum critical points yield the expressions~(\ref{ecoherente}) which give the minimum energy of the system together with information about the quantum phase transition present in the Dicke Hamiltonian. 

At the phase transition the expectation values of the number of photons, the number of atoms in excited states and their fluctuations exhibit a sudden increase from zero. It was shown that these properties scale with the number of atoms, that their reported divergences at the critical point represent the limit when this number goes to infinity, and that in this limit they remain divergent in the superradiant phase. This implies that no singular behavior in the number of photons, the number of excited atoms, their fluctuations, squeezing and quadratures will be found around the phase transition in future experiments. The presence of entanglement \cite{Lam04} and chaos \cite{Ema03a} at the phase transition in the Dicke model should also be analyzed from this perspective.

This work was partially supported by CONACyT-M\'exico, FONCICYT (project-94142), and DGAPA-UNAM .


\begin{thebibliography}{1}

\bibitem{dicke} 
R. H. Dicke, Phys. Rev. {\bf 93}, 99 (1954).

\bibitem{hepp}
K. Hepp and E. H. Lieb, Phys. Rev. {\bf A 8}, 2517 (1973).

\bibitem{eberly}
J. H. Eberly, N. B. Narozhny, and J. J. Sanchez-Mondragon,  Phys. Rev. Lett. {\bf 44}, 1323 (1980).

\bibitem{kulinski}
J. R. Kuklinski, and J.L. Madajczyk, Phys. Rev. {\bf A 37}, 3175 (1988).

\bibitem{baumann}
K. Bauman, C. Guerlin, F. Brennecke, and T. Esslinger, Nature {\bf 464}, 1301 (2010).

\bibitem{Nat10}
P. Nataf and C. Ciuti, Nature Comm. {\bf 1}:72 (2010).

\bibitem{Dim07}
F. Dimer, B. Estienne, A.S. Parkins, and H.J. Carmichael, Phys. Rev. {\bf A 75}, 013804 (2007).

\bibitem{nagy}
D. Nagy, G. Konya, G. Szirmai, and P. Domokos, Phys. Rev. Lett. {\bf 104}, 130401 (2010).

\bibitem{Ema03}
C. Emary and T. Brandes, Phys. Rev. {\bf E 67}, 066203 (2003).

\bibitem{Cas10} O. Casta\~nos, E. Nahmad-Achar, R. L\'opez-Pe\~na, J. G. Hirsch,
{\it Symmetries in Nature, Symposium in Memoriam Marcos Moshinsky}, AIP Conf. Proc. {\bf 1323} (2010) 40 - 51.

\bibitem{ocasta1}
O. Casta\~nos, R. L\'opez-Pe\~na, E. Nahmad-Achar, J.G. Hirsch, E. L\'opez-Moreno, and J. E. Vitela, Phys. Scr. {\bf 79 }, 065405 (2009).

\bibitem{manko}
{\sl Theory of Nonclassical States of Light}, Eds. V.V. Dodonov and V.I. Man'ko (Taylor \& Francis, London, 2003).

\bibitem{meystre}
P. Meystre and M. Sargent III, {\sl Elements of Optics}
(Springer, New York,1991).

\bibitem{gilmore1972}
F.T. Arecchi, E. Courtens, R. Gilmore, and H. Thomas, Phys. Rev. 
{\bf A6}, 2211 (1972).

\bibitem{Cas05} O. Casta\~nos, R. L\'opez-Pe\~na, J. G. Hirsch, and E. L\'opez-Moreno.
Phys. Rev. {\bf B 72} (2005) 012406,

\bibitem{gilmore3}
R. Gilmore and L.M. Narducci, Phys. Rev. A {\bf 17}, 1747 (1978);
R. Gilmore, {\sl Catastrophe Theory for Scientists and Engineers},
(Wiley, New York, 1981).

\bibitem{ocasta2}
O. Casta\~nos, E. Nahmad-Achar, R. L\'opez-Pe\~na, and J.G. Hirsch, Phys. Scr. {\bf 80 }, 055401 (2009).

\bibitem{Cas06}
O. Casta\~nos, R. L\'opez-Pe\~na, J. G. Hirsch, and E. L\'opez-Moreno,
Phys. Rev. {\bf B 74} (2006) 104118.

\bibitem{Ema03a} C. Emary and T. Brandes, Phys. Rev. Lett. {\bf 90}, 044101 (2003).

\bibitem{Chen08} G. Chen, X. Wang, J.-Q. Liang, Z.D. Wang, Phys. Rev. {A \bf 78}, 023634 (2008).

\bibitem{Kita93}  M. Kitagawa and M. Ueda, Phys. Rev. A {\bf 47}, 5138 (1993).

\bibitem{Lam04} N. Lambert, C. Emary and T. Brandes, Phys. Rev. Lett. {\bf 92}, 073602 (2004).




 






\end{thebibliography}
\end{document}